\documentstyle[sprocl]{article}
\input{psfig}
\def\be{\begin{equation}}
\def\ee{\end{equation}}
\def\bea{\begin{eqnarray}}
\def\eea{\end{eqnarray}}

\begin{document}
\title{AN EFFICIENT MULTIPROCESSOR MANAGEMENT SYSTEM FOR EVENT--PARALLEL 
COMPUTING}
\vspace*{-1.25in}
\centerline{Proceedings Preprint, \, DPF\,'96, \, Minneapolis, Minnesota  
 (10--15 August 1996)}
\vspace*{0.75in}
\author{ DON SUMMERS, STEVE BRACKER, KRISHNASWAMY GOUNDER,\\ KEVIN HENDRIX }
%\address{Department of Physics and Astronomy, University of 
%Mississippi--Oxford,\\ University, MS 38677, USA}
\address{Dept.~of Physics and Astronomy, University of 
Mississippi, Oxford, MS 38677}
\maketitle\abstracts{
Performance of software using TCP/IP sockets to distribute events to UNIX
workstations is described. This simple software was written at the University
of Mississippi to control UMiss farm reconstruction of 8 billion raw events,
part of Fermilab E791's data.  E791 reconstructed HEP's largest data set to
study charm physics.} 
\vspace*{-3mm}

Fermilab E791 wrote a big dataset (50 Terabytes, 20 billion
events, 24{\thinspace}000 8mm Exabyte tapes) in 1991 and early 
1992.\footnotemark{\,}\footnotemark{\,}\footnotemark {} Reconstruction 
challenged available computing,
requiring over 10$^4$ mips-years. The task was larger
than at colliders (Table~1). Reconstruction
was nevertheless completed using four {\em farm} 
sites.\,\footnotemark {} Here we describe the multiprocessor management 
software\,\footnotemark {} developed 
and run at the University of Mississippi farm
(Figs.~1 and 2 show hardware).

HEP events are usually independent.
Interprocess I/O isn't needed. 
An efficient parallel system {\em just} has to input and output
events fast enough so {\em clients} are never idle. 
Management software had to do a lot of hard work
in early HEP 
systems.\footnotemark\footnotemark {} Clients had minimal operating systems. 
All data had to be formatted in 
a server and downloaded into clients word by word. Moving 
from \hfill single \hfill to \hfill multiple \hfill CPUs \hfill was \hfill hard;
\hfill the \hfill division \hfill between \hfill server \hfill and 
\vspace*{-1.5mm}
\begin{figure}[!bhtp] 
\psfig{figure=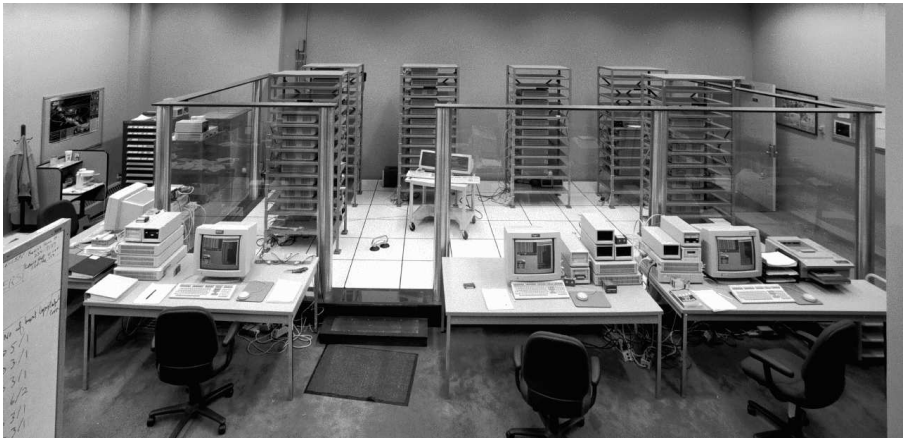,height=2.275in} 
\vspace*{-3mm}
\caption{Mississippi farm overview.  Servers are on the 
four tables.  Clients are on the racks shown 
and on desktops not shown.  
The espresso machine 
is on the left.}
\end{figure}
\begin{flushright}\begin{minipage}[t]{2.6in}
client code was intricate.
With the advent of {\em commercial workstation} 
clients\,\,\footnotemark {} with real
operating systems, most work inherent in moving
to multiprocessors vanished. Using
Network File System software, \hfill server  
\end{minipage}
\begin{minipage}[t]{85mm}
\vspace*{2pt}
disks can be {\em cross-mounted} so that files are accessible by multiple
clients. In this model, 
even inexpensive
diskless clients directly read an executable code
file, a run number file, calibration files, a raw input record file of events,
and write report files and reconstructed event files.  The server writes
input events from tape to disk files. At the end of a job, the server
copies client output event files to tape and combines client reports, as
clients work on the next job. 
Because 85\% of E791 events were filtered away after
reconstruction, disk output was fast enough for us.  Event input by disk
also worked, but too slowly. So, our multiprocessor manager bypasses disk
for input using instead 
Transmission Control Protocol/
Internet \hfill 
Protocol.\footnotemark {} With \hfill TCP/IP, \hfill 
processes \hfill make \hfill a   
\end{minipage} 
\begin{minipage}[t]{77mm}
connection between themselves and pass
data back and forth using {\em read\_from\_connection} and
{\em write\_through\_connection} subroutine calls. 
A test of TCP/IP gave 900 kbyte/s, ending
client idleness.
Fig.~3 illustrates how the network I/O calls are used. As the server prepares
to start a client, it uses {\em make\_socket} to ``have a phone put in'', so
that it will be able to connect to the client. 
When the client starts, it too
uses {\em make\_socket} to ``have a phone put in''. The server ``lists its
number'' by binding its socket to a port ({\em bind\_socket}), and ``stays near
the phone'' listening for an attempt to connect ({\em listen\_socket}). The
client ``calls up'' the server ({\em connect\_socket}) and the server ``picks
up the phone'' establishing the connection ({\em accept\_socket}). 
When it needs input data, the client ``places its order'' by writing a message
to the server ({\em write\_socket}). The server is continually monitoring all
of the client connections for requests  
({\em select\_socket}). \hfill When \hfill a \hfill request \hfill comes 
\hfill in, \hfill the 
\end{minipage}\end{flushright} 

\begin{figure} [!htbp]
\vspace*{-6.60in}
\psfig{figure=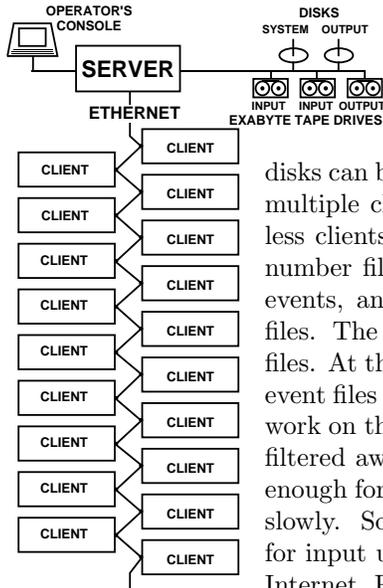,height=3.0in} 
\parbox{39mm}{\caption{Mississippi farm architecture.
Servers and clients are DECstation 5000 workstations running ULTRIX.
Some have MIPS R3000 CPUs; others the more powerful 
MIPS R4000. Altogether, there are 68 processors organized into
4 farms isolated by Ethernet bridges.  One typical farm is shown here.
The two input tape drives alternate automatically. Each drive is only wearing 
itself out half the time. 
If an input tape drive fails, the next tape starts automatically.
The output is staged through disk and streamed to tape.
If the output tape drive fails output data can easily be recovered from disk.
If a disk fills, processing is automatically paused until space {\em appears}.
This I/O scheme avoids constant operator supervision.}}
\end{figure}
\begin{flushright}\begin{minipage}[t]{77mm}
\vspace*{-4.7mm}
server ``writes down the order''  ({\em read\_socket}), \hfill and
\end{minipage}\end{flushright} 

\begin{table}[thbp]
\begin{flushright}
\vspace*{-2mm}
\parbox{48mm}{\caption{
E791 raw data size    
and $p \overline{p}$,  
$e^- p$, \ and $e^+ e^-$ collider experiment sizes. D0 saves digitized 
waveforms.}}
\vskip 3pt
{\footnotesize \tabcolsep=0.7mm
\renewcommand{\arraystretch}{1.15}
\begin{tabular}{|lccc|} \hline
Exper-            & Events         & Tera-      & Recording \\
iment              & $ \div 10^6$   & bytes      & Period  \\ \hline 
E791     & 20\,000 & 50     &  7/91 - 1/92  \\ 
CDF      & 95          & 10     & 10/85 - 12/95 \\ 
D0       & 80          & 40     &  2/92 - 12/95 \\ 
H1       & 70          & 2.5     &  5/92 - 12/95 \\ 
ZEUS     & 50          & 5     &  5/92 - 12/95 \\ 
Aleph     & 60          & 1.7   &  8/89 - 11/95 \\ 
Delphi    & $\sim$30    & $\sim$5 &  8/89 - 11/95 \\ 
L3        & 83          & 3.4     &  8/89 - 11/95 \\ 
OPAL      & 102         & 1.5     &  8/89 - 11/95 \\ 
CLEO     & 600         & 5     & 10/79 - 12/95 \\ \hline
\end{tabular}} \end{flushright} \end{table} 
\begin{flushright}\begin{minipage}[t]{45mm}
\vspace*{-9.5mm}
does its best to satisfy the client's request. The client and server shuttle
messages back and forth (each writing to the other and reading from the other)
until the input  
data is exhausted. Next,
the server notifies the
clients that \hfill there \hfill is \hfill no \hfill more \hfill data.
\end{minipage}\end{flushright} 

\begin{figure}[!htbp]
\vspace*{-100.2mm}
\psfig{figure=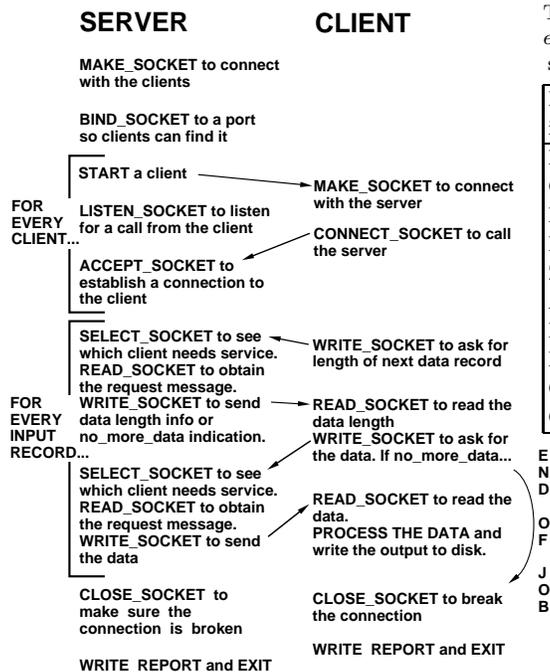,height=3.5in} 
\vspace*{-0.1in}
\parbox{62mm}{\caption{TCP/IP socket communication.}}
\end{figure}

\begin{flushleft}\begin{minipage}[t]{4.7in}
\vspace*{-1.2mm}
The clients then finish their tasks, close
their connections ({\em close\_socket}), and exit; the server finishes its
tasks and exits. 
The Fortran-callable C routines that manipulate sockets and connections really
are that simple to use. Only {\em read\_socket} is more than
a C to Fortran interface, and even its trivial. Most of
the real work has already been done in
UNIX, TCP/IP, and Berkeley Sockets. 
\end{minipage}\end{flushleft}

Although most of our farm processors are in racks, some
are on people's desks. We have found
it satisfactory to allow users to abruptly kill the client process whenever
they find its activities on their workstations to be troublesome. A
reconstruction code crash also kills a client.  In either case, the server
is quickly aware of the dead client and adjusts event
distribution. A disadvantage of this approach is that a few
input events are {\em trapped} and lost. 
Having 20 billion events, we
take a
rather cavalier attitude.  In E791, processing raw events is rather
like hauling corn to market in a truck. If a few grains of corn fall out
of the truck, no one cares. The alternative approach -- treating events as
babies in a hospital nursery, where one normally expects a somewhat stricter
accounting -- only makes sense later with small selections of {\em interesting}
events. 

Before writing our own multiprocessor software we considered
extracting the few features needed from large packages under development
at Fermilab\,\,\footnotemark{\,}\footnotemark {} and 
Argonne.\footnotemark {} However, offsite
support was unavailable.  
So we focused on writing software which could do
a limited number of things very well; {e.g.} run many clients per 
server
efficiently and tolerate client crashes and operating
system upgrades. 
Six man--weeks
were spent coding. Farm
operation required 5 hours over a day.
After seeing our approach to moving farm 
data, the D0 experiment decided to follow a
similar strategy.\footnotemark

Funding in June 1993 allowed an expansion of the UMiss farm
from 1100 to 2900 mips. By July 1993, the increased computing
was acquired and processing data. E791 reconstruction was completed in
Sept.~1994. A total of 8 billion events on 10{\thinspace}000 raw data
tapes were processed in Mississippi. Before running final
reconstruction, dozens of {\em full farm} tests of algorithms for actual
charm yield were run, each test for a few days. The charm yield
tripled. 
X Window operator control displays written in Tcl/Tk aided bookkeeping.
Tape reading was multiply buffered, so that events were almost always
available immediately when a client asked for them. During smooth running,
timing CPUs showed that at least 97\% of client processing cycles were
used. Overall efficiency, considering cycles lost for {\em any} reason,
exceeded 90\% over a 2$\frac{1}{2}$ year period. 

Efficient management of multiple processors has led to the
reconstruction of 200{\thinspace}000 charm particles, the world's
largest sample. 
Results\,\,\footnotemark {} include
DPF\,'96 papers by N.~Copty, L.~Cremaldi, K.~Gounder, M.~Purohit, K.C.~Peng,
A.~Tripathi, R.~Zaliznyak, and C.~Zhang.
We especially thank Lucien Cremaldi and Breese Quinn for their contributions to
building and running the UMiss farm. This work was supported in part by
U.S.~DOE DE-FG05-91ER40622. 
\vspace*{-5mm}
\footnotetext[1]{S.~Amato {\em et al.,} {\em Nucl.~Instr.~Meth.}
{\bf A324} (1993) 535; \ E791 DA.}
\footnotetext[2]{D.J.~Summers 
{\em et al.,} XXVIIth Rencontre de Moriond, 
Les Arcs (15-22 March 1992) 417.}
\footnotetext[3]{B.R.~Kumar, 
{\em Vertex Detectors,} Plenum Press, Erice (21-26 September 1986) 167.}
\footnotetext[4]{CBPF--Rio de Janeiro, Fermilab, 
Kansas State, Mississippi{\,}(CREMF).}
\footnotetext[5]{Steve Bracker {\em et al., 
IEEE Trans.\ Nucl.\ Sci.} {\bf NS-43} (October 1996).}
\footnotetext[6]{Paul F.\ Kunz {\em et al.,} 
{\em IEEE Trans.\ Nucl.\  Sci.} {\bf NS-27} (1980) 582; 
\ IBM 168 emulator.}
\footnotetext[7]{J.~Biel {\em et al.,} 
{\em Computer Physics Communications} {\bf 45} (1987) 331; \ ACP.}
\footnotetext[8]{C.\ Stoughton and D.J.\ Summers, 
{\em Computers in Physics} {\bf 6} (1992) 371.}
\footnotetext[9]{Sidnie Feit, 
{\em TCP/IP: Architecture, Protocols, and Implementation}, McGraw-Hill, 1993.}
\footnotetext[10]{F.~Rinaldo and S.~Wolbers, 
{\em Computers in Physics} {\bf 7} (1993) 184.}
\footnotetext[11]{Aleardo Manacero,
CPS Performance under Different Network Loads,
Fermi{\,}Pub-94-33.}
\footnotetext[12]{{R.J.} Harrison, 
{\em International Journal of Quantum Chemistry} {\bf 40} (1991) 847;
\ TCGMSG.}
\footnotetext[13]{Kirill Denisenko {\em et al.,} D0 Farm Production System, 
CHEP\,'94, 152; \ Fermilab $p \overline{p}.$}
\footnotetext[14]{{E.M.} Aitala {\em et al.} (E791 Collaboration), 
Observation of D--$\pi$ Production Correlations
in 500 GeV $\pi^-$--N Interactions 
(submitted to {\em PRL}); \
Search for $D^0$--$\overline{D}{}{^0}$ Mixing in 
Semileptonic Decays, {\em Phys.~Rev.~Lett.} {\bf 77} (1996) 2384; \
Mass Splitting and Production of 
$\Sigma_{\hbox{c}}^0$ and $\Sigma_{\hbox{c}}^{++}$ Measured in 
500 GeV/c $\pi^-$--N Interactions,  
{\em Phys.~Lett{.}} {\bf B379} (1996) 292; \
Asymmetries Between the
Production of $D^+$ and $D^-$ Mesons from 500 GeV/c $\pi^-$--Nucleus
Interactions as a Function of $x_F$ and $p_t^2$, 
{\em Phys.~Lett{.}} {\bf B371} (1996) 157; \
Search for the Flavor-Changing Neutral Current 
Decays
$D^+ \negthinspace \negthinspace \rightarrow \negthinspace 
\pi^+ \mu^+ \mu^-$ and $D^+ 
\negthinspace \negthinspace \rightarrow \negthinspace 
\pi^+ e^+ e^- \negthinspace ,$
{\em Phys.~Rev.~Lett{.}} {\bf 76} (1996) 364.}
%
%
%\begin{figure} \rule{5cm}{0.2mm}\hfill\rule{5cm}{0.2mm} 
%\vskip 2.5cm \rule{5cm}{0.2mm}\hfill\rule{5cm}{0.2mm}
%\psfig{figure=filename.ps,height=1.5in} \caption{Radiative (off-shell, off-page and out-to-lunch) SUSY Higglets. \label{fig:radish}}
%\end{figure}
%
\end{document}